\documentclass[sigconf,screen]{acmart}
\acmDOI{10.1145/3663529.3663842}
\acmYear{2024}
\copyrightyear{2024}
\acmSubmissionID{fsecomp24industry-p46-p}
\acmISBN{979-8-4007-0658-5/24/07}
\acmConference[FSE Companion '24]{Companion Proceedings of the 32nd ACM International Conference on the Foundations of Software Engineering}{July 15--19, 2024}{Porto de Galinhas, Brazil}
\acmBooktitle{Companion Proceedings of the 32nd ACM International Conference on the Foundations of Software Engineering (FSE Companion '24), July 15--19, 2024, Porto de Galinhas, Brazil}
\received{2024-02-08}
\received[accepted]{2024-04-18}
\newcommand{\leak}{{\small{\textsf{mrDetector}}}\xspace}

\AtBeginDocument{%
  \providecommand\BibTeX{{%
    \normalfont B\kern-0.5em{\scshape i\kern-0.25em b}\kern-0.8em\TeX}}}

\usepackage[ruled, linesnumbered, commentsnumbered, noend]{algorithm2e}
\usepackage{balance}
\usepackage{graphicx}
\usepackage{caption}
\usepackage{enumitem}
\usepackage{subfigure}
\usepackage{makecell}
\usepackage{multirow}
\usepackage{threeparttable}
\usepackage{xcolor}
\usepackage{booktabs}
\usepackage{url}

\begin{document}

\title{Combating Missed Recalls in E-commerce Search: A CoT-Prompting Testing Approach}

\author{Shengnan Wu}
\orcid{0000-0003-1964-313X}
\affiliation{%
  \institution{School of Computer Science, Fudan University}
  \city{Shanghai}
  \country{China}
}
\email{snwu19@fudan.edu.cn}

\author{Yongxiang Hu}
\orcid{0009-0003-5099-2335}
\affiliation{%
  \institution{School of Computer Science, Fudan University}
  \city{Shanghai}
  \country{China}
}
\email{yongxianghu23@m.fudan.edu.cn}

\author{Yingchuan Wang}
\orcid{0009-0001-0767-1662}
\affiliation{%
  \institution{School of Computer Science, Fudan University}
  \city{Shanghai}
  \country{China}
}
\email{yingchuanwang23@m.fudan.edu.cn}

\author{Jiazhen Gu}
\orcid{0000-0002-5831-9474}
\affiliation{%
  \institution{School of Computer Science, Fudan University}
  \city{Shanghai}
  \country{China}
}
\email{jiazhengu@cuhk.edu.hk}

\author{Jin Meng}
\orcid{0009-0008-7037-977X}
\affiliation{%
  \institution{Meituan}
  \city{Beijing}
  \country{China}
}
\email{mengjin02@meituan.com}

\author{Liujie Fan}
\orcid{0009-0007-7319-6904}
\affiliation{%
  \institution{Meituan}
  \city{Beijing}
  \country{China}
}
\email{fanliujie@meituan.com}

\author{Zhongshi Luan}
\orcid{0009-0007-0852-115X}
\affiliation{%
  \institution{Meituan}
  \city{Beijing}
  \country{China}
}
\email{luanzhongshi@meituan.com}

\author{Xin Wang}
\orcid{0000-0002-9405-4485}
\affiliation{%
  \institution{School of Computer Science, Fudan University}
  \city{Shanghai}
  \country{China}
}
\email{xinw@fudan.edu.cn}

\author{Yangfan Zhou}
\orcid{0000-0002-9184-7383}
\affiliation{%
  \institution{School of Computer Science, Fudan University}
  \city{Shanghai}
  \country{China}
}
\email{zyf@fudan.edu.cn}

\renewcommand{\shortauthors}{S. Wu, Y. Hu, Y. Wang, J. Gu, J. Meng, L. Fan, Z. Luan, X. Wang, and Y. Zhou}
\begin{abstract}
Search components in e-commerce apps, often complex AI-based systems, are prone to bugs that can lead to missed recalls—situations where items that should be listed in search results aren't. This can frustrate shop owners and harm the app's profitability.
    However, testing for missed recalls is challenging due to difficulties in generating user-aligned test cases and the absence of oracles. In this paper, we introduce \leak, the first automatic testing approach specifically for missed recalls. To tackle the test case generation challenge, we use findings from how users construct queries during searching
     to create a CoT prompt to generate user-aligned queries by LLM. In addition, we learn from users who create multiple queries for one shop and compare search results, and provide a test oracle through a metamorphic relation. Extensive experiments using open access data demonstrate that \leak outperforms all baselines with the lowest false positive ratio. Experiments with real industrial data show that \leak discovers over one hundred missed recalls with only 17 false positives. 
\end{abstract}



\maketitle

\section{introduction}\label{sec:introduction}
Search components play an important role in modern e-commerce platforms ({\em e.g.}, Amazon~\cite{luo2022query} and Yelp~\cite{payne2021powering}).
In such platforms, shop owners present products/services their shops provide to customers, who discover products/services by searching on the platforms.
The search components of such platforms are usually complicated AI-based retrieval systems that take user-generated queries as input and search results, \textit{i.e.} lists of entries (typically shops or products), as output. Apart from the user-generated queries, user preferences and searching contexts also affect the search results. Bridging users, shop owners and products for sale together, those search components cast a non-negligible impact on the revenue-making and success of e-commerce apps~\cite{degenhardt2021report}. 

Complication is a notorious cause of bugs ~\cite{lyu1996handbook}. As complicated AI-based retrieval systems, search components in e-commerce apps are not bug-free. Existing work mainly focuses on false recalls,  {\em i.e.}, the situations where an entry appears in the search result but should not be recalled according to algorithmic and business logic~\cite{vaughan2004new, hannak2013measuring, ganguly2015word, zhou2015metamorphic, van2017remedies, nigam2019semantic, zhang2020towards, huang2020embedding,li2021embedding}. Those false recalls cause search results to be unstable, irrelevant, and inconsistent with user preferences. However, little research attention has been paid to situations where an entry should be recalled according to algorithmic and business logic but does not appear in the search result ({\em i.e.}, the {\em missed} recalls). Missed recalls bring quite negative impacts. For example, customers may be dismayed to see their favorite diner neglected by the search component. Shop owners may be dissatisfied to see their shops not being presented to customers. As a result, they may decide to switch to competitor platforms, and eventually, the platform's profitability will suffer.

In real industrial settings, missed recalls are not rare. 
Take \textit{M-App}, a prevalent e-commerce app developed by Meituan, one of the largest online shopping provides 
as an example.
In the second quarter of 2023, over 20 percent of problematic search results are confirmed to be associated with missed recalls. Currently, missed recalls are typically found by feedback from shop owners 
. This post-mortem way of missed-recall-discovery has two major drawbacks. Firstly, it heavily relies on manual efforts and can only discover a limited number of missed recalls. Secondly, it is accompanied by complaints from shop owners who have found their shops can not be displayed to customers normally. Hence, an approach to detect missed recalls automatically and proactively is necessary.

Nevertheless, it is non-trivial to test a search component automatically and proactively targeting missed recalls. Two major challenges remain unsolved. Firstly, it is far from easy to generate \textbf{realistic} test cases automatically (challenge 1). In this scenario, test cases are user-generated natural language queries. How users express their shopping needs and the search context, which are subjective and ambiguous, should be involved in the generation process. Without realistic test cases, we will only end up with defects induced by edge cases that users seldom input into a search component in an e-commerce app. In the meantime, the \textbf{oracle} problem awaits (challenge 2). Unlike false recalls that stand out in the search results naturally, missed recalls are more salient and user-oriented. Specifically, it is the users, instead of the e-commerce app that hold the ultimate standards in judging missed recalls. Those user standards are subjective, perceptual and may vary among different users. Hence, given a specific case, it is accomplishable by human effort to judge whether there is a missed recall. However, it is challenging to formulate the way of human judgment into a set of automatic algorithms. 

The user-oriented nature of missed recalls inspires us to analyze historical track of missed recalls, learn from the handling procedure and design our approach accordingly. We present missed recall Detector (\leak), which relies on a metamorphic relation to provide a test oracle and leverages the power of the large Language Model (LLM) to generate realistic test cases. 
The rationale behind this is that subjective and ambiguous domain knowledge about interaction and need expression are implicitly obtained by LLMs via a large training corpus. We also use the findings from analyzing historical missed recalls to guide prompt engineering and ensure context-compatible test cases. 
To our knowledge, \leak is the first testing approach targeting missed recalls in search components of e-commerce apps. It can also be generalized to other e-commerce apps besides \textit{M-App}.


Extensive evaluation is conducted to compare the performance of \leak in different settings against multiple baselines. Experiments with open data show that \leak outperforms all baseline methods with the lowest false positive ratio. 
The ablation study revealed that an LLM with a larger parameter size, the customized chain of thought (CoT) prompt mimicking how general users construct queries and an LLM based validation step benefit the performance of \leak. 
We also deployed \leak in the field to detect real-world online missed recalls for \textit{M-App}.
In the real-world experience, we have detected over one hundred missed recalls with only 17 false positives. 



The contributions of this paper are summarized as follows:
\begin{itemize}[leftmargin=*]
\item 
To the best of our knowledge, we are the first to conduct a comprehensive study into missed recalls of search components in the e-commerce setting, which can shed light on future work of evaluating and optimizing complicated AI-based search components. 
\item 
 We present \leak, a testing approach targeting missed recalls of search components in e-commerce apps. It leverages LLM to overcome the challenge of realistic test cases and a metamorphic relation to overcome the challenge of the oracle problem. It provides references to testing work involving the subjectivity of human and faced with the oracle problem.
\item 
Experiments with open data illustrate that each major component benefits the performance of \leak, outperforming all baseline methods with the lowest false positive ratio. Experiments with industrial data show the handiness of \leak in the real industrial setting, with over 100 missed recalls found and over half of them estimated reproducible. 
\end{itemize}

\section{Background and Motivation}
\label{section: The Motivating Example}

The AI-based search components play an important role. 
According to the annual report\footnote{\url{https://media-meituan.todayir.com/202403221654401765350700_tc.pdf}}
of Meituan, the \textit{M-App} has over 600 million active users and has established business cooperation with over 9 million shops. As users rely on the search component to discover what to buy ~\cite{sondhi2018taxonomy} and shops rely on the search component to complete orders, bugs in search components not only degrade user experience, but also hinder the profitability of shop owners and the e-commerce apps themselves.

However, as complex AI-based retrieval systems, search components are prone to bugs. Search components take multiple factors into account to retrieve products or shops according to user-constructed queries. Prominent ones include semantic similarity between the query and the target, the geographic location of the target, the time of the search, and the corporation's business strategies. To enhance user experience, user preferences and click-through history are also considered, which adds to the complexity. Each step in this complicated retrieval process may introduce bugs, including ones incurring missed recalls. 

\begin{figure*}[ht]
    \centering
    \subfigure[The search result of query ``Chen's hardware''.]{
        \begin{minipage}[t]{0.255\linewidth}
        \centering
        \includegraphics[width=\linewidth]{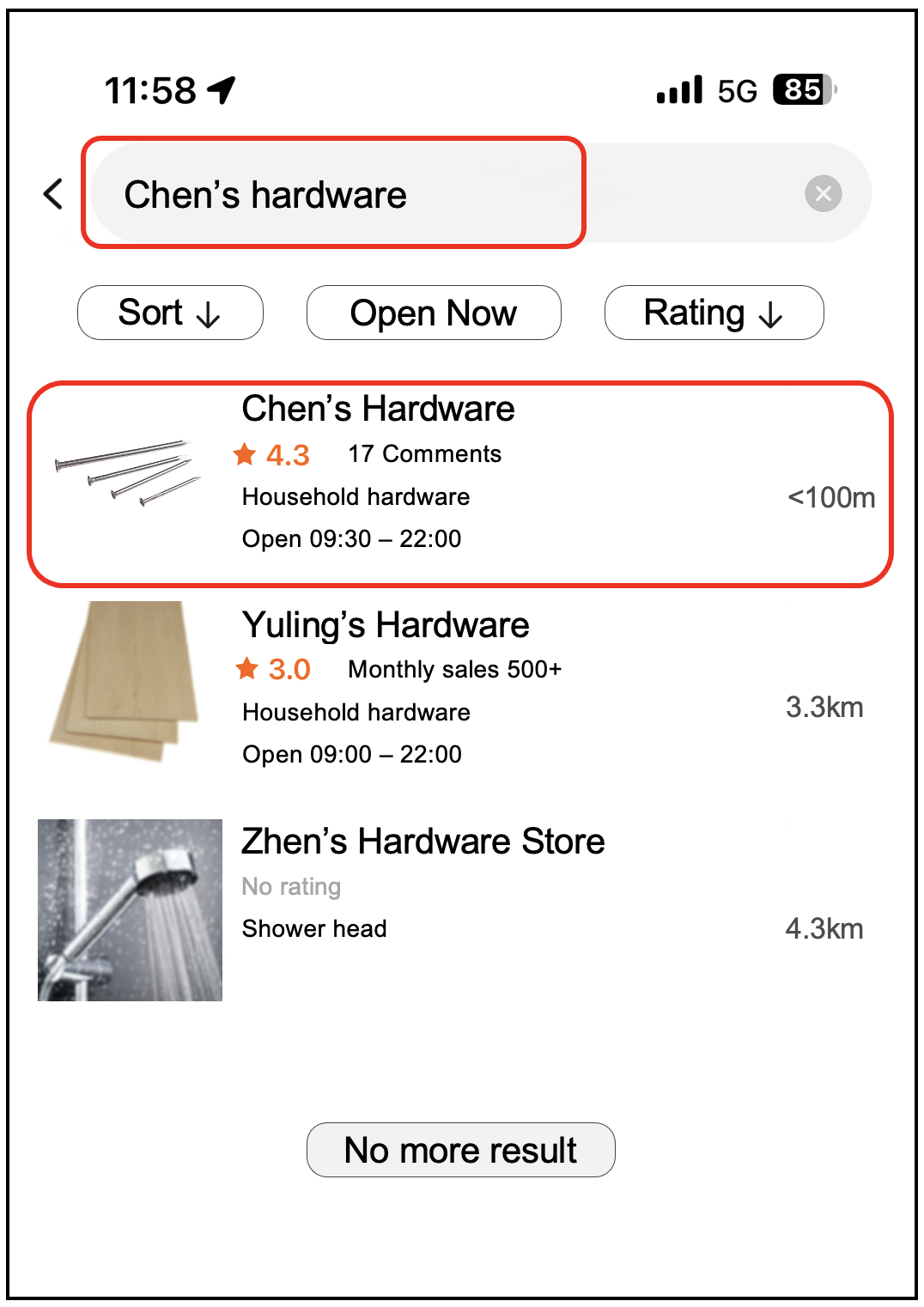}
        \vspace{-4pt}
        \label{fig:case_1}
        \end{minipage}
    }
    \subfigure[The search result of query ``Chen's''.]{
        \begin{minipage}[t]{0.255\linewidth}
        \centering
        \includegraphics[width=\linewidth]{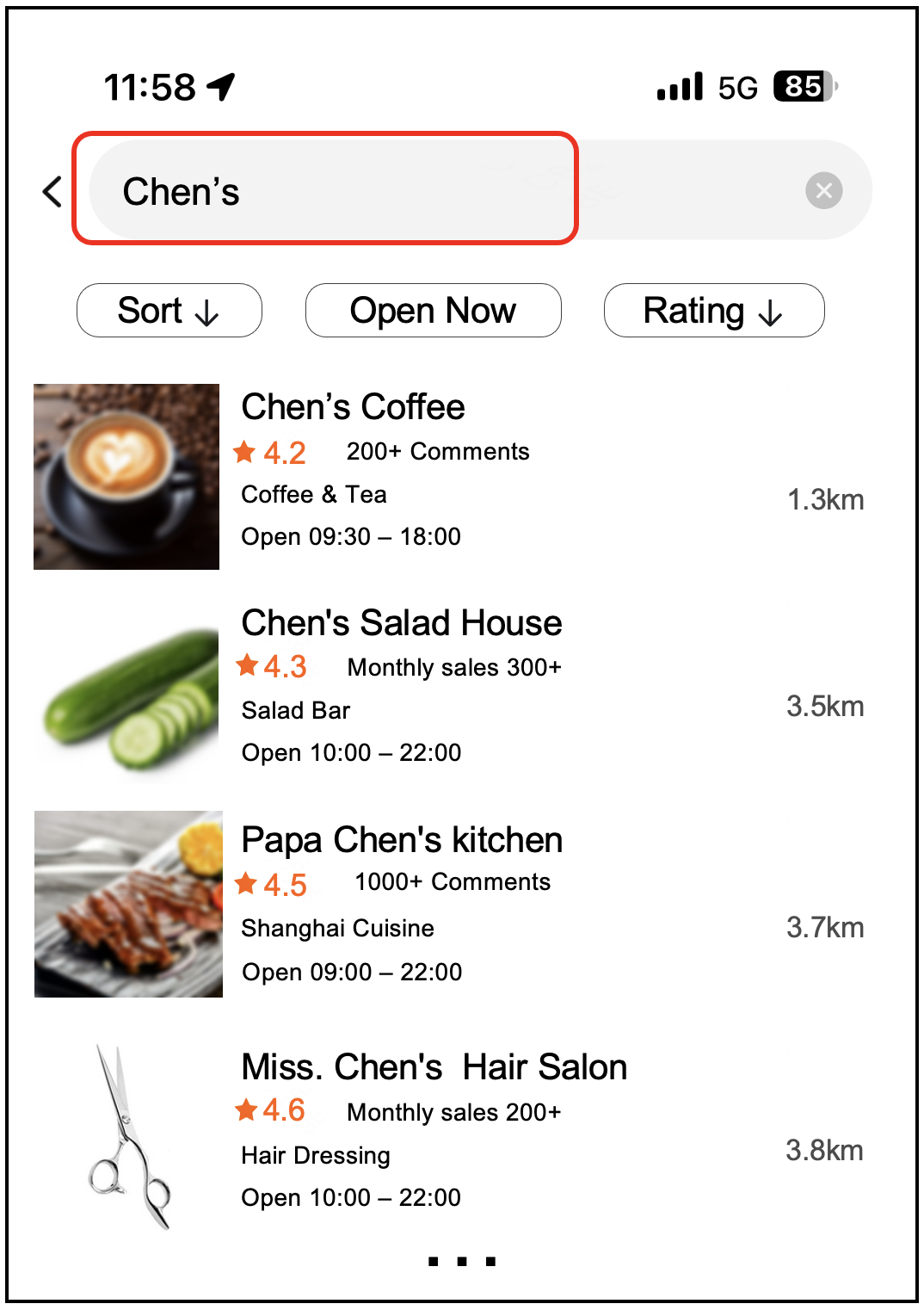}
        \vspace{-4pt}
        \label{fig:case_2}
        \end{minipage}
    }
    \subfigure[The user reported ticket describing a missed recall of ``Chen's'']{
        \begin{minipage}[t]{0.276\linewidth}
        \centering
        \includegraphics[width=\linewidth]{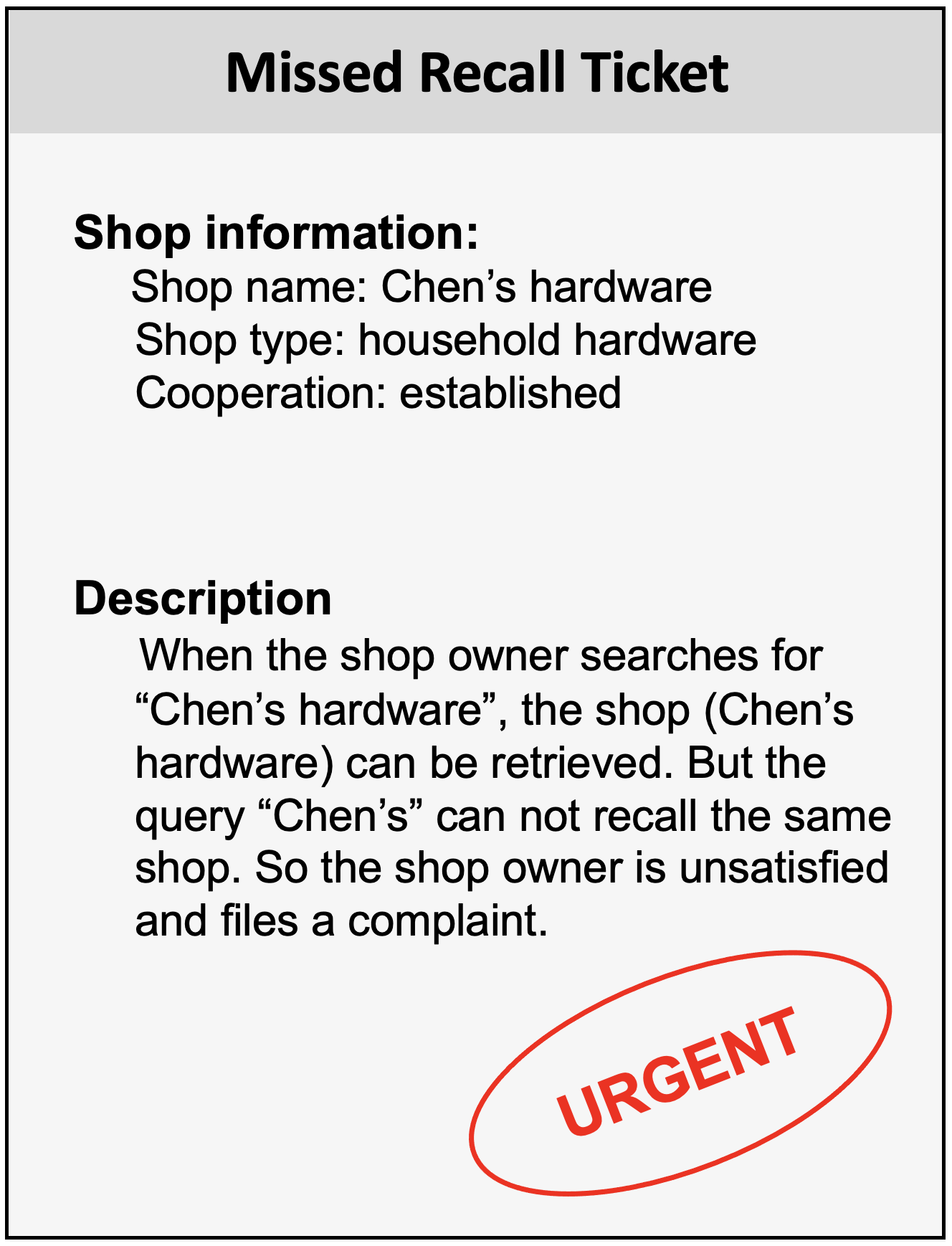}
        \vspace{-4pt}
        \label{fig:case_3}
        \end{minipage}
    }
    \vspace{-12pt}
    \caption{An example of  missed recall}
    \label{fig:example}
    \vspace{-8pt}
\end{figure*}

Missed recalls are estimated to be the second-largest reason for problematic search results in the second quarter of 2023. Figure~\ref{fig:example} demonstrates a missed recall reported by the shop owner. For privacy concerns, the information presented is desensitized. As presented in Figure~\ref{fig:example}, the shop owner initiates two searches targeting his own shop. The target shop can be retrieved when he uses the shop's full name as the query. But it cannot get recalled when he searches for \textit{Chen's}. 

Currently, missed recalls are mainly discovered by feedback from shop owners and confirmed heavily relying on manual efforts. To confirm a missed recall in any e-commerce app typically takes four significant steps. Engineers first examine whether the query inducing a missed recall conveys search intention towards the target shop. Then, engineers make sure the location of the search is not too far away from the location of the target shop, as shops far away possess lower priority during retrieval. Next, engineers examine the time of the search. If the target shop is searched out of its opening time, it takes lower priority during retrieval. Finally, engineers consult business operators to check whether the target shop has violated some business policies. If so, the factor of violation punishment may explain this missed recall. After several rounds of communication involving engineers, business operators and the shop owner, the missed recall is confirmed. \textit{Chen's} coveys search intention towards the target shop \textit{Chen's hardware} and \textit{Chen's} is aligned with how general users search on e-commerce apps~\cite{ai2017learning,ai2019zero}.The shop owner searches for his shop exactly in that shop and during the opening time. This target shop has no record of policy violation. The missed recall presented is confirmed not a false positive.

However, this back-and-forth communication in confirming usually takes days, leaving missed recalls continuing to compromise shop owners' satisfaction. Due to dissatisfaction, the shop owner files a complaint as presented in Figure~\ref{fig:example}. 
The long confirming process substantiates the challenging nature of testing towards missed recalls. Subjectivity of search intention expression presents challenges for test case generation, for test cases are user queries conveying search intentions. Diverse and complicated confirming standards can not be easily automated, which adds to the absence of an oracle, as those standards involve manually trading off features of the concerning search component, business strategies and shop status.

\section{An Empirical Study on Historical Missed Recalls}
\label{section: The Empirical Study}
Given the challenging nature of testing towards missed recalls due to the subjective test cases and absence of an oracle, we first conduct an empirical study based on historical missed recalls. This empirical study provides insights on how general users discover missed recalls and hence inspire the method design.

Specifically, we took user-reported missed recalls handled during January 2023 to June 2023 at Meituan as the study object. There are around 100 entries in total\footnote{We purposely omit the actual number for protection of the operational status of the corporation.} 
. We qualitatively analyze them by semantic coding~\cite{terry2017thematic}. Two authors independently coded all user-reported missed recalls. New codes are created until no new information emerges. After the coding process, the two authors solved differences by discussing and consulting corresponding engineers from the corporation.

We found that the standards used in confirming missed recalls are complicated, detailed, and somewhat ambiguous, which can hardly be automated, proving the challenge of lacking an oracle. Moreover, users typically provide multiple examples, \textit{i.e.} queries for the same target shop, and compare the search results when reporting missed recalls. This finding also motivates us to consider providing an oracle by an metamorphic relation. Here we present the complicated standards in confirming and how users manage to identify missed recalls despite the complicated standards. For privacy concerns, the shop and query information presented is desensitized.

\subsection{Ambiguous and Complicated Standards}\label{section:complicated standard}

Both subjective and objective factors are considered to confirm a missed recall. More importantly, the factors considered vary according to different missed recalls. Here we summarize several significant factors.
\begin{itemize}[leftmargin=*]
\item 
\textbf{Reasonableness of the query:} As the search component is designed for human usage, confirmed missed recalls can only be induced by \textit{human-perceively reasonable} queries. In practice, a query should align with how general users express shopping needs and convey explicit intention towards the target shop. For example, if a shop named \textit{lovely pets} located 100 meters east of \textit{Becker street post office} can not be retrieved by the query\textit{100 meters east of Becker street post office}, it is not a missed recall. General users usually do not express their shopping needs towards a pet store in that fashion. However, how shopping intentions are expressed involves the subjectivity of human nature, which can hardly be encoded into definite rules.
\item 
\textbf{The geography:} Distance affects search results. Shops far away have lower priority during retrieval at the \textit{M-App}. Hence, long distances may cause false positives of missed recall. 

\begin{figure*}[ht!]
    \centering
    \includegraphics[width=0.9\linewidth]{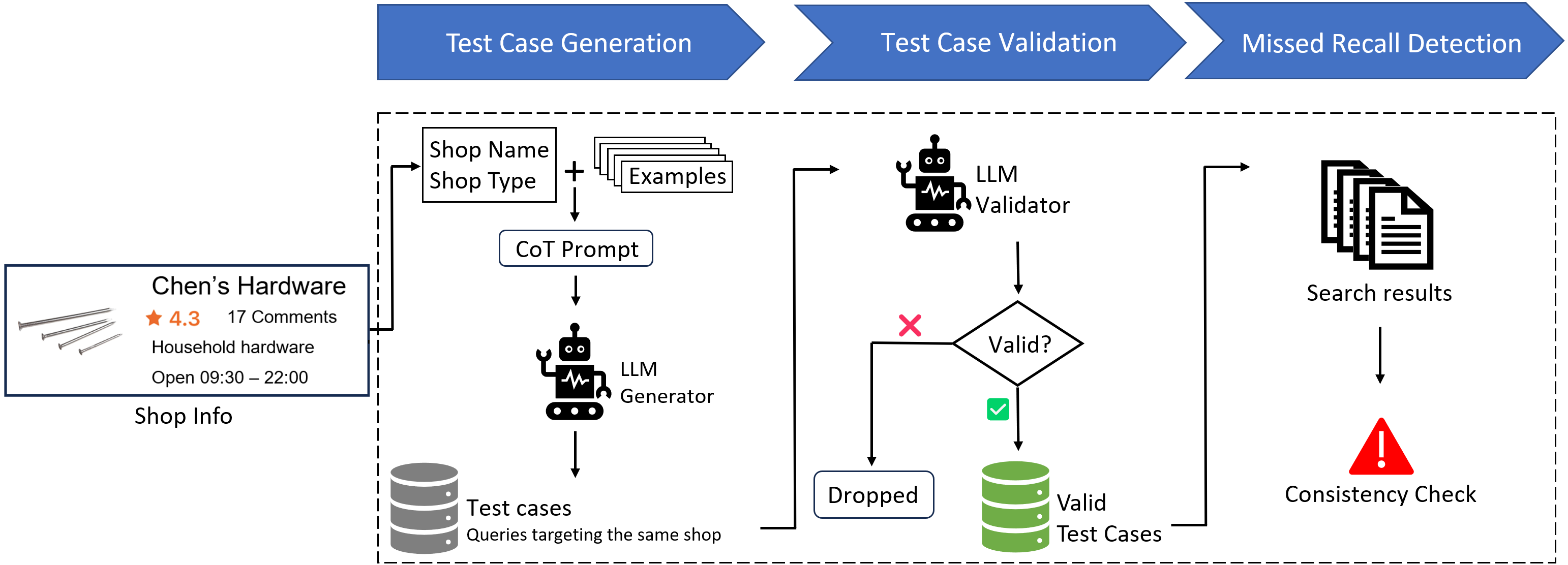}
    \vspace{-8pt}
    \caption{Overview of \leak}
    \label{fig:overview}
    \vspace{-8pt}
\end{figure*}

\item 
\textbf{Operating status:} 
Usually, only open shops will be retrieved by most e-commerce apps. If a certain shop is closed when a user initiates a search towards it in the \textit{M-App}, it is normal that it can not be retrieved. 
\item 
\textbf{Ordering strategies:} 
To enhance user experience, e-commerce apps incorporate multiple ordering strategies, affecting the search results. For example, if a user searches for \textit{hotpot}, those shops offering discounts due to sales promotion may be prioritized during retrieval. Given the fixed number of shops exposed to users, a false impression of missed recall may be presented. To give another example, administrative punishment, lowering the priority in retrieval, works similarly. 
\item 
\textbf{Others:} User preferences and click-through history of the user affect search results as well. Those factors may also create false impressions of missed recalls.
\end{itemize}

The above findings demonstrate the complication of identifying missed recall and further prove the oracle problem faced by testing towards missed recalls automatically and proactively. However, despite the complicated standards, general users manage to report missed recalls, with limited knowledge about the operational strategy of the corporation. Hence, we present how general users manage to identify missed recalls and use these findings to inspire our method design.

\subsection{Multiple Queries for the Same Shop}
\label{multiple queries}

On reporting missed recalls, users typically compare the search results of multiple queries towards the same target shop. Accounting for all confirmed user-reported missed recalls (around 80\footnote{We purposely omit the actual number for protection of the operational status of the corporation.}\label{fn:omit}), about 3 queries are constructed for the same shop averagely. Without much knowledge about the confirmation standards used in the corporation, users report missed recalls only when those queries return inconsistent search results. For example, 3 queries are constructed for one target shop, and two queries can recall this shop but one can not. Then, this user reports a missed recall.

After analyzing all queries involved in the user-reported missed recalls we study (over 200 queries \footnotemark[\value{footnote}]), we obtain 2 observations. We observe user queries are diverse and prone to the personal style of expression, which rules can hardly mimic, but what users consider when constructing queries remains stable.

To illustrate the subjectivity and diversity of user queries, we demonstrate the following three types of queries as examples.
\begin{itemize}[leftmargin=*]
\item 
\textbf{Equivalent queries.} In concept, the queries are basically the equivalents of the shop name. For example, \textit{hardware Chen's} is an equivalent query for the shop~\textit{Chen's hardware}. 
\item 
\textbf{Including queries.} Including queries conceptually include the target shop. For example, \textit{Indian Restaurant} is an including query for the shop \textit{Ali's curry house}. 
\item 
\textbf{Included queries.} Included queries are conceptually included by the target shop. For example, \textit{spicy hotpots} is an included query for a restaurant serving spicy and non-spicy hotpots. 
\end{itemize}

Unlike user queries, diverse and subjective, what users consider when constructing queries remains relatively stable. Here we present prominent information types users consider when constructing queries.
\begin{itemize}[leftmargin=*]
\item 
\textbf{Shop name.} Users usually take the full name of the shop, characters from the full name of the shop and initials of the shop as queries. For example, given a target shop \textit{Ma's burgers}, users may also search for \textit{Ma's} and \textit{hamburgers}.
\item 
\textbf{Products/ services the shop offers.} Users also search for a certain shop by the products or services it offers, like \textit{hotpot}, \textit{haircut} and \textit{pedicures}. 
\item
\textbf{Geographic location of the shop.} Users may avoid typing the shop's exact name due to cognitive workload. So they search vaguely by geographic location and construct queries like \textit{hotpot People's square} and \textit{dumplings nearby}.
\end{itemize}

In summary, findings from the reported missed recalls substantiate the challenges of testing targeting missed recalls and user practices also hint at solutions. Constructing multiple queries towards the same shop and comparing the search results inspires us to consider metamorphic testing to overcome the oracle problem. Subjectivity of human nature leads to diverse queries, which inspires us not to rely on pure rules, but to incorporate subjective factors of humans in test case generation. As users typically consider shop names, services and products the shop provides and the location of the shop when constructing queries, we can use shop names and shop types as input data for test case generation. For shop names and shop types include the three types of information.
\section{mrDetector approach}
\label{section: methodology}
This section presents the technical design of our proposed approach \leak. It is an automatic testing approach targeting missed recalls in the search results. Inspired by the findings from section ~\ref{section: The Empirical Study}, we leverage the ability of LLM to conduct test case generation, which solves the challenge of test cases in line with how general users express their shopping needs. Inspired by the finding that users identify missed recalls by constructing multiple queries for one target shop and comparing the search results, we provide an oracle by a metamorphic relation, which solves the challenge of lacking oracles.

The overall workflow of \leak is illustrated in figure~\ref{fig:overview}. \leak includes three steps: 1) LLM-based test case generation, 2) test case validation, and 3) missed recall detection. 

\begin{figure*}[ht!]
    \centering
    \includegraphics[width=0.95\linewidth]{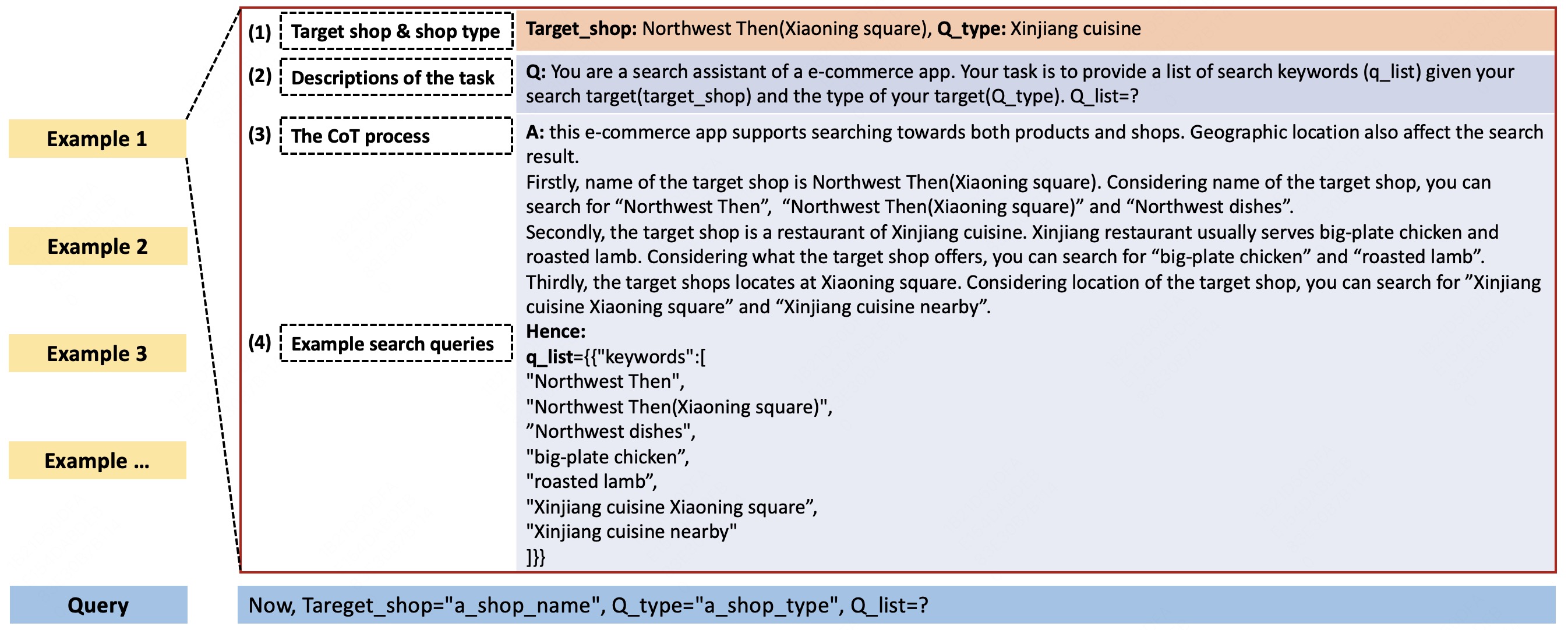}
    \vspace{-8pt}
    \caption{A simplified demonstration of the prompt we use to generate test cases}
    \label{fig:prompt}
    \vspace{-8pt}
\end{figure*}

In the LLM-based test case generation step, \leak generates user queries aligning with what general user search with on e-commerce apps, as test cases. In particular, a customized CoT prompt, which mimics how general users construct queries, is used. However, due to the inherent hallucinations in LLMs, test cases not inline with how general users search on e-commerce apps are inevitable. In the test case validation step, the LLM is called again to rethink on and judge the reasonableness of the generated test cases. This step can identify unreasonable test cases that traditional techniques like semantical similarity can not identify and hence reduce false positives. In the missed recall detection step, missed recalls are detected by violation of a metamorphic relation.


\subsection{Test Case Generation}
In this step we generate test cases/queries automatically by the LLM. As we use a metamorphic relation to provide test oracle, our test cases (essentially user queries) must: (1) be real enough, approximating to user-constructed ones; (2) support the metamorphic relation. To generate real test cases, we leverage the ability of the LLM instead of traditional NLP techniques. LLMs are trained with a large corpus of natural languages, implicitly conveying how humans express their shopping needs and interact with search components. It is reasonable to assume that given the information needed, the LLM could generate queries that approximate what users search with on the search component. 
After trading off the costs and base abilities, we choose GPT-3.5 turbo for query generation. To support the metamorphic relation, we generate a group of test cases/queries for one shop instead of a single test case/query. Those queries all target the same shop and express search intention towards it.  

We tailor the prompt to mimic how general users construct queries during their search behavior. According to what information general users consider when constructing queries, we take shop names and shop types as the input. The shop names also include information about where the shop is located. To leverage the ability of LLM to generate queries close to what general users construct during searching, we use a series of techniques, including in-context learning\cite{min2022rethinking}, CoT\cite{zhang2022automatic}, question-answering examples~\cite{brown2020language} and human in the loop\cite{ge2023openagi}. 

We first break the query generation process of general users (illustrated in section ~\ref{multiple queries}) into three steps (shop name based generation, services/products based generation and location based generation)
according to the information types they consider. Then we accordingly implement the three steps into the prompt with the CoT technique. Next we use the in-context learning technique to provide domain knowledge about our specific task and include multiple examples in the prompt. As LLMs benefit from examples in a question-answer format in the prompt~\cite{brown2020language}, all examples are provided in the question-answer format. To ensure the quality of all the examples in the prompt, we randomly sample from real user queries of Meituan and manually revise them to achieve consistency in our scenario. We also use human evaluation to substitute all examples that general users think "\textit{not real enough}" due to revision. Finally, we use the human-in-the-loop technique and iteratively improve our prompt based on the human feedback on the quality of the queries generated. We stop the iteration process until no unreal ones are found in randomly sampled 50 generated queries. Our final prompt consists of four parts: (1) target shops and shop types; (2) descriptions of the task; (3) the CoT process of query generation; (4) example search queries. The prompt is written in Chinese, as the input data are also in Chinese. Figure ~\ref{fig:prompt} provides a simplified demonstration of our prompt. For privacy concerns, the shop information and example queries presented are desensitized.

\subsection{Test Case Validation}

Due to hallucination issues, LLMs may generate unreliable content despite the customized CoT prompt to ensure generation quality. For example, the query \textit{a barber} is generated for the target shop, a hairdressing salon. Another example is \textit{supermarket near A} for the target shop \textit{A supermarket}, which confuses A as a location. 

Those test cases are human-perceively unreasonable but can not be detected by traditional techniques. For example, the \textit{a barber} scenarios would achieve high semantical similarity scores and the \textit{supermarket near A} scenarios would achieve high textual matching scores.

Hence, we leverage a rethink scheme~\cite{liu2023make} to reduce false positives induced by unreasonable test cases in the test case validation step. Specifically, given the name and type of the target shop and the generated test case, we call the LLM again to re-think the generated test cases and drop ones that do not align with how general users express their shopping needs. The rationale is that multiple trials with the LLM decrease the probability of unreliable responses incurred by hallucination. Only test cases validated as \textbf{resonanble} are kept for the next step. The prompt used for test case validation follows the few-shot strategy. Multiple examples, both reasonable test cases to be kept and unreasonable test cases to be dropped, are included in the prompt. 

\subsection{Missed Recall Detection}
In this step, we determine whether a certain test case incurs a missed recall by consistency checks. We propose the metamorphic relation that test cases generated for one target shop should present consistent search results. Violations of this metamorphic relation indicate missed recalls. Hence, the test oracle is formally expressed below.

Given test cases \( X_1, X_2, \ldots, X_n \) for a target shop $t$, the function \( R(X) \) represent search results recalled by $X$ . Formally, for any \( X_n \), we define \( y_n \) as follows:
\begin{equation}
y_n = \begin{cases} 
\text{True} & \text{if } t \in R(X_n), \\
\text{False} & \text{if } t \notin R(X_n).
\end{cases}
\end{equation}
The oracle suggests a potential missed-recall when, for any pair \( i, j \) where \( 1 \leq i, j \leq n \), we have \( y_i \neq y_j \).


We report test cases that do not recall the target shops as those incurring missed recalls only when other test cases with the same target shop can recall it. To prevent false positives, we are particularly cautious about the situation that all test cases towards a specific shop can not recall the target shop. For example, if a shop decides to cease its business cooperation with an e-commerce app, whatever user queries you try on the search component of that app can not recall this shop. Those situations should not report missed recalls.

      

In implementation, an internal test API that provides the same functionalities to the search component is used to get all search results. As shown in section ~\ref{section:complicated standard}, apart from the query, many other factors also affect the search results of search components in e-commerce apps. To prevent false positives, we use the same user account to ensure the same user profiling and change the location of searches to the accurate geographic locations of the target shops to prevent the explainability of long distances. We also conduct searches during the time slot of 10 am. to 9 pm., when most shops are open, which prevents the explainability of shop status.


\section{evaluation}\label{section:Experiments}
In this section, we evaluate the performance of \leak by answering the following research questions (RQs):
 
\begin{itemize}[leftmargin=*]
    \item \textbf{RQ1}: How do different LLMs affect the performance of \leak?
    \item \textbf{RQ2}: How do different strategies of prompt engineering affect the performance of \leak?
    \item \textbf{RQ3}: How does the LLM validation step affect the performance of \leak?
    \item \textbf{RQ4}: How does \leak perform in detecting real online missed recalls?
\end{itemize}

RQ1, RQ2 and RQ3 target the effect of each part in \leak, 
\textit{i.e.}, the LLM used, the prompt for test case generation and the validation step, on its performance. RQ4 treats \leak as a whole, and aims to illustrate its handiness in detecting real online missed recalls of a prevalent e-commerce app in China. 

\subsection{Experimental Setup}
\textbf{Datasets.} We use two datasets to evaluate the performance of \leak.
\begin{itemize}[leftmargin=*]
    \item \textbf{Dataset A}: Dataset A is a manually constructed open-access dataset consisting of 600 entries of shops. The shop entries are collected via the Baidu Map API\footnote{\url{https://api.map.baidu.com/place/v2/search}}, which returns nearby shops given a geographic location. We randomly sampled these shops from all shops in Beijing and Shanghai via the API in November 2023. Those 600 entries cover most daily life services, including hairdressing, skin care, nail polishment and catering. We believe such diverse shop types could unveil the ability of \leak to detect general missed recalls, avoiding the possibility that \leak overfits on a certain shop type.
    \item \textbf{Dataset B}: 600 entries of shops randomly sampled from the business partner table of Meituan. This table records all shops that have established business cooperation with Meituan. The shop types are in line with that of Dataset A. Dataset B is a dataset of real business data. 
\end{itemize}
Dataset A and Dataset B employ the same data format and are all in Chinese. As shown in Table~\ref{tab:data_example}, we demonstrate how each shop is presented in each entry with a mock shop.
\begin{table}
\centering
\caption{How to present a shop in our datasets and the purpose of each attribute.}
\vspace{-8pt}
\label{tab:data_example}
\resizebox{0.5\textwidth}{!}{
\begin{tabular}{lllll}
\hline
\textbf{\makecell[l]{Shop name}} & \textbf{\makecell[l]{Shop type}} & \textbf{\makecell[l]{City}} & \textbf{\makecell[l]{Latitude}} & \textbf{\makecell[l]{Longitude}} \\
\hline
\makecell[l]{Old Flavor Hotpot} & \makecell[l]{Beijing hotpot} & \makecell[l]{Beijing} & \makecell[l]{116.3E} & \makecell[l]{40.5N} \\
\hline
\multicolumn{2}{c|}{\textbf{\makecell[l]{Input data for test case \\generation}}} & \multicolumn{3}{c}{\textbf{\makecell[c]{Detailed location to \\reduce test false positives}}} \\
\hline
\end{tabular}
}
\vspace{-8pt}
\end{table}


\textbf{Metrics}.
We use Reported Cases ($N_{reported}$), Confirmed Cases ($N_{Confirmed}$), False Positive Ratio
($R_{fp}$), and Test Case Efficiency ($E_{tc}$) to measure the performance of \leak. 

\begin{table*}[t!]
\centering
  \caption{Reported Cases ($N_{reported}$), Confirmed Cases ($N_{Confirmed}$), False Positive Ratio ($R_{fp}$), and Test Case Efficiency ($E_{tc}$) comparison among \leak versions implemented with LLMs of different parameter sizes.}
  \vspace{-8pt}
  \label{tab:RQ1}
  \renewcommand{\arraystretch}{1}
  \resizebox{.8\textwidth}{!}{
  \begin{tabular}{cc|cc|cc}
    \toprule
      \multicolumn{2}{c}{\textbf{Versions}}&\multicolumn{2}{c}{\textbf{Benefit metics}}&\multicolumn{2}{c}{\textbf{Cost metrics}}\\ 
      LLMs&Parameters&$N_{reported}$$\uparrow$&$N_{Confirmed}$$\uparrow$&$R_{fp}$$\downarrow$&$E_{tc}$$\downarrow$\\
    \midrule
      GPT-Neo&2.6B&35 entries/ 35 shops&6 entries/6 shops&29/35=0.829&2607/6=434.500\\
      Chat-GLM2&6B&54 entries/ 44 shops&32 entries/26 shops&22/54=0.407&3803/32=118.844\\
      Qwen&14B&\textbf{64 entries/ 48 shops}&\textbf{54 entries/40 shops}&10/54=0.156&4375/54=81.019\\
      GPT-3.5 turbo&over 100B&47 entries/ 33 shops&46 entries/33 shops&\textbf{1/47=0.021}&\textbf{3724/46=80.95}\\
   \bottomrule
  \end{tabular}}
  \vspace{-8pt}
\end{table*}

$N_{Reported}$ and $N_{Confirmed}$ describe the \textbf{benefit} side of \leak. $N_{reported}$ measures how many missed recalls are reported by \leak. Higher $N_{reported}$ indicates stronger ability to discover missed recalls. $N_{Confirmed}$ measures how many reported missed recalls are confirmed by human. Higher $N_{Confirmed}$ indicates higher accuracy of \leak. We present $N_{Reported}$ and  $N_{Confirmed}$ both entry-wise and shop-wise. $R_{fp}$ and $E_{tc}$ describe the \textbf{cost} side of \leak, and are calculated as:
\begin{equation}
  R_{fp}=\frac{N_{Reported}-N_{Confirmed}}{N_{Reported}}  
\end{equation}
\begin{equation}
  E_{tc}=\frac{N_{Total}}{N_{confirmed}} 
\end{equation}
where $N_{Total}$ refers to the total number of test cases/queries during one round of test. Lower $R_{fp}$ indicates less human efforts incurred by false positives in confirming. $E_{tc}$ measures the average number of test cases needed to find a confirmed missed recall. Lower $E_{tc}$ hints a smaller number of test cases needed to find the same amount of missed recalls and hence a higher test efficiency. $R_{fp}$ and $E_{tc}$ are reported entry-wise.

\textbf{Experimental environment.}
We implement \leak with 612 lines of Python codes. All experiments are conducted on an Ubuntu 20.04 server with Intel (R) Xeon (R) Platinum 8175M CPU (2.50GHz), 48GB RAM and two NVIDIA GeForce RTX 4090 GPUs (210MHz).

\textbf{Manual Confirmation Procedure.} Manual confirmation of reported missed recalls requires examining the reasonableness of test cases, which is prone to the subjectivity of the conductor. During confirmation, two authors first independently confirm reported missed recalls. Then, they solve differences by discussing and consulting a third party. We believe such a cross-validating procedure curbs the effects of human subjectivity on confirmation results.

\subsection{Performance with Different LLMs (RQ1)}

\textbf{Baselines.} \leak use GPT-3.5 turbo, which is estimated to have over 100B parameters~\cite{ye2023comprehensive}, to conduct test case generation. Here we compare the performances of \leak with LLMs of different parameter sizes to unveil how the selection of the LLM affects the performance. Besides \leak, we implement three baseline methods with three LLMs of different parameter sizes.
\begin{itemize}[leftmargin=*]
    \item \textbf{GPT-Neo Version}: \leak implemented with GPT-Neo. GPT-Neo~\cite{gpt-neo} is a pre-trained language model following the transformer architecture. According to the official document, it has around 2.6B parameters.
    \item \textbf{ChatGLM2 Version}: \leak implemented with ChatGLM2. ChatGLM2 is an open bilingual language model based on General Language Model framework~\cite{du2022glm}. It leverages both the Chinese and English languages and has approximately 6B parameters. 
    \item \textbf{Qwen Version}: \leak implemented with Qwen 14-B. It ~\cite{qwen} is a 14B-parameter version of the large language model series, Qwen (abbr. \textit{Tongyi Qianwen}), proposed by \textit{Alibaba Cloud}. 
\end{itemize}

\textbf{Results.} We answer RQ1 with Dataset A. As demonstrated in Table ~\ref{tab:RQ1}, comprehensively \leak performs best with GPT-3.5 turbo, which has the largest parameter size.
Specifically, on the cost side \leak only reports one false positive and only needs 80.95 test cases to discover a confirmed missed recall with GPT-3.5 turbo. This achieves the best among all LLMs concerned. On the benefit side, \leak built on GPT-3.5 turbo shows a comparable ability with that built on Qwen, which performs best with benefit metrics.

According to the results in Table~\ref{tab:RQ1}, a conclusion could be loosely drawn that larger parameters benefit the performance of \leak. Analysis with false positives also confirms this conclusion. Test cases not in line with what general users search with on e-commerce apps are the most significant reason for incurring false positives. Chat-NEO, which has the minimal-sized
parameters, often outputs random characters or emojis that hardly convey search intentions toward the target shops. Chat-GLM2 and Qwen do not output random characters but sometimes fail to catch the subjective way general users express shopping needs via user queries. For example, in the Chinese language context, a \textit{seafood joint} usually refers to where sells raw seafood. While a \textit{seafood restaurant} instead refers to a restaurant that sells seafood dishes. So general users seldom use \textit{seafood joint} to search for a seafood restaurant on e-commerce apps. 
GPT-3.5 turbo, which has the maximal-size parameters, only generates one test case that is not real enough. 
Analysis with false positives may also partially explain why \leak build on Qwen performs best with benefit metrics. Qwen includes a larger ratio of Chinese corpus in the training data. Hence, it may enjoy advantages in dealing with Chinese input data.
\begin{center}
\noindent\fbox{
    \parbox{0.9\linewidth}{\textbf{Summary:} A larger parameter size generally facilitates the ability to generate valid test cases of \leak. Comprehensively \leak built on GPT-3.5 turbo achieves the best performance. It reports 46 confirmed missed recalls with only one false positive.}
}
\end{center}

\subsection{Performance with Different Prompt Engineering Strategies (RQ2)}
\textbf{Baselines.} \leak uses a customized CoT prompt to generate test cases that both align with users' search behavior and support our metamorphic relation. Here we investigate how different prompt engineering strategies affect the performance of \leak. Besides \leak, we implement two baseline methods with two different prompt engineering strategies.
\begin{itemize}[leftmargin=*]
    \item \textbf{Few-shot Version}: \leak implemented with a standard few-shot prompt and GPT-3.5 turbo. The standard few-shot prompt includes all examples in the customized CoT prompt, but no hints about the \textit{steps} of test case generation are provided. 
    \item \textbf{Zero-shot Version}: \leak implemented with a standard zero-shot prompt and GPT-3.5 turbo. Only textual descriptions of the generation task are included in the zero-shot prompt. No examples and hints about the \textit{steps} of test case generation are provided.   
\end{itemize}

\begin{table*}[h!]
\centering
  \caption{Reported Cases
  ($N_{reported}$), Confirmed Cases ($N_{Confirmed}$), False Positive Ratio ($R_{fp}$), and Test Case Efficiency ($E_{tc}$) comparison among \leak versions implemented with different prompt engineering strategies.}
  \vspace{-8pt}
  \label{tab:RQ2}
  \renewcommand{\arraystretch}{1}
  \resizebox{.9\textwidth}{!}{
  \begin{tabular}{cc|cc|cc}
    \toprule
      \multicolumn{2}{c}{\textbf{Versions}}&\multicolumn{2}{c}{\textbf{Benefit metics}}&\multicolumn{2}{c}{\textbf{Cost metrics}}\\ 
      Prompts&info.&$N_{reported}$$\uparrow$&$N_{Confirmed}$$\uparrow$&$R_{fp}$$\downarrow$&$E_{tc}$$\downarrow$\\
    \midrule
      Few-shot&examples&\textbf{156 entries/ 98 shops}&\textbf{121 entries/76 shops}&35/156=0.223&\textbf{5700/121=47.107}\\
      Zero-shot&-&46 entries/ 42 shops&39 entries/37 shops&7/46=0.152&4622/39=118.513\\
      Customized CoT&Examples, steps&47 entries/ 33 shops&46 entries/33 shops&\textbf{1/47=0.021}&3724/46=80.95\\
   \bottomrule
  \end{tabular}}
\end{table*}

\begin{table*}
\centering
  \caption{Reported Cases
  ($N_{reported}$), Confirmed Cases ($N_{Confirmed}$), False Positive Ratio ($R_{fp}$), and Test Case Efficiency ($E_{tc}$) comparison among \leak versions implemented with/without the LLM validation step.}
  \vspace{-8pt}
  \label{tab:RQcheck}
  \renewcommand{\arraystretch}{1}
  \resizebox{.78\textwidth}{!}{
  \begin{tabular}{c|cc|cc}
    \toprule
      \multirow{2}{*}{\textbf{Versions}}&\multicolumn{2}{c}{\textbf{Benefit metics}}&\multicolumn{2}{c}{\textbf{Cost metrics}}\\ 
      ~&$N_{reported}$$\uparrow$&$N_{Confirmed}$$\uparrow$&$R_{fp}$$\downarrow$&$E_{tc}$$\downarrow$\\
    \midrule
      without LLM validation&\textbf{78 entries/ 45 shops}&\textbf{71 entries/42 shops}&7/78=0.090&\textbf{3724/71=52.451}\\
      with LLM Validation&47 entries/ 33 shops&46 entries/33 shops&\textbf{1/47=0.021}&3724/46=80.95\\
   \bottomrule
  \end{tabular}}
  \vspace{-8pt}
\end{table*}

\textbf{Results.} We answer RQ2 with Dataset A. As demonstrated in Table ~\ref{tab:RQ2}, comprehensively \leak performs best with the customized CoT prompt. \leak implemented with a standard few-shot prompt generates far more test cases than that implemented with the zero-shot prompt and customized CoT prompt. 
Hence, it is reasonable that it reports the most missed recalls. Judging from the benefit side, the few-shot version discovers around three times more missed recalls than the customized CoT prompt version. 
However, judging from the cost side, it reports over 30 times more false positives than the customized CoT prompt version. Trading off costs and benefits, comprehensively, the customized CoT prompt achieves the best performance with its significantly low $R_{fp}$.

We develop two assumptions based on the above observation and analysis with test cases generated. Firstly, examples in the prompt may increase the number of test cases generated. Examples are materials that the LLM can learn from and mimic, hence may inspire more generation results. This partially explains why the few-shot version generates the most test cases. Secondly, the steps stating how a generation task should be solved and contained in the customized CoT prompt may guide how the LLM \textit{thinks}. So, how the LLM generates test cases may approximate how general users construct queries, and hence less test cases, especially unreal test cases, are generated. This assumption partially explains why the customized CoT prompt version archives the lowest $R_{fp}$, six times lower than the second lowest. 

\begin{center}
\noindent\fbox{
    \parbox{0.9\linewidth}{\textbf{Summary:} 
     Comprehensively \leak built on the customized CoT prompt performs best with its lowest $R_{fp}$, six times lower than the second lowest. Firstly, examples included in the prompt may increase the number of test cases generated. Secondly, the steps stating how a generation task should be solved and contained in the customized CoT prompt may guide how the LLM \textit{thinks} and reduce false positives.}
}
\end{center}

\subsection{Performance with/without the LLM Validation (RQ3)}
\textbf{Baseline.} To reduce the impact of hallucination issues, \leak leverage a rethink scheme and call the LLM again to validate the generated test cases. To illustrate how effective this LLM validation step is, we implement the \textbf{Without LLM Validation Version} of \leak as the baseline. Except for omitting the LLM validation step, no further changes are made.

\textbf{Results.} We answer RQ3 with Dataset A. As shown in Table~\ref{tab:RQcheck}, the LLM validation step significantly reduces the false positive ratio (3.28 times). ALL false positives the validation step prevents are due to edge cases not aligning with how general users search on e-commerce apps. For example, \textit{Shanghai [space] restaurant} is generated for a restaurant serving Shanghai Cuisine. However, in the Chinese context, this test case is generally interpreted as \textit{restaurants in Shanghai}, instead of \textit{Shanghai Cuisine Restaurants}. This edge case, which can not be easily detected by traditional techniques like semantic similarity and textual mating, proves an LLM validation step is necessary.


This validation step with LLM helps reduce the count of missed recalls that were reported and confirmed due to our cautious approach in the experiment. When the LLM cannot provide a clear decision for test cases, we consider them as failing the LLM check step. This experimental approach reduces the number of test cases and consequently lowers the count of reported missed recalls.
\vspace{-2pt}
\begin{center}
\noindent\fbox{
    \parbox{0.9\linewidth}{\textbf{Summary:} 
     The LLM validation step brings down the false positive ratio 3.28 times, only compromising 35 percent of confirmed missed recalls. It can prevent false positives incurred by edge cases not aligning with how general users search on e-commerce apps and can not easily detected by traditional techniques. }
}
\end{center}

\subsection{Performance in Discovering Real Online Missed Recalls (RQ4)}


To show how well \leak works in real industries, we use Dataset B, which contains actual industry data. This data, handled by different departments, is often untidy. While doing well with open data can demonstrate algorithmic strengths, it doesn't guarantee success in real industrial scenarios, where results are affected by both algorithms and other factors.

\textbf{Results.} In total, 6396 test cases are generated by \leak. 118 entries of missed recalls (involving 91 shops) are reported. After human confirmation, 101 entries (involving 76 shops) are confirmed. The $R_{fp}$ is 0.144, which means false positives take 14 percent of human efforts during confirming. The $E_{tc}$ reaches 63.327. This demonstrates that on average 63 test cases are needed to find a confirmed missed recall. 


Revisiting the results of RQ1 to RQ3, we find \leak generates more test cases and finds more missed recalls with industrial data. One possible explanation is that the customized CoT prompt is inspired by the historical missed recalls, which are based on real industrial data. This may benefit the performance of \leak in a real industrial setting. 

We also find that \leak reports more false positives with real industrial data. During confirmation, we find those false positives are incurred not only by unreal test cases (the same as in RQ1, RQ2, and RQ3) but also by the quality of the input data. Although not common, it exists that the shop type provided is reasonable but too broad to describe a shop from the subjective aspects of humans. For example, a shop curing nail fungus often is not considered a \textit{nail polish center} by general users, although it does deal with nails. 


Effective as \leak is in detecting online missed recalls, according to engineers in charge, it can further help reduce future missed recalls if it reports missed recalls that are \textbf{reproducible} on the \textbf{phone}. To prepare fixes, the first step for engineers is to reproduce those missed recalls on the phone and analyze their impact on the user experience. Hence, although reproducibility does not compromise the effectiveness, it increases the handiness of \leak in a real industrial setting. To this end, we randomly sample 15 confirmed missed recalls and manually reproduce them on a mobile phone. Specifically, we search for the same test cases/user queries on the \textit{M-App} installed on an iPhone 13, and check whether the same missed recalls appear. During the searches, we roughly change the geographic location by filling in the nearest landmark buildings to the target shops. Over half (8/15) of missed recalls can be reproduced. After analyzing those that can not be reproduced, we find the most prominent reason lies in the outdatedness of the target shop information. Some shops no longer exist but they haven't been erased from the business partner table. So, they are taken as target shops during the test. This explains 4 out of 7 missed recalls that can not be reproduced. This reproduction experiment further illustrates the handiness of \leak, besides its effectiveness in detecting missed recalls.
\begin{center}
\noindent\fbox{
    \parbox{0.9\linewidth}{\textbf{Summary:} 
     \leak finds 118 missed recalls with real industrial data. Only 17 are confirmed to be false positives. Over half of the missed recalls confirmed are estimated to be reproducible. The customized CoT prompt of \leak is inspired by the historical missed recalls, which are based on real industrial data. This may benefit the performance of \leak in a real industrial setting. }
}
\end{center}
\section{Case Studies}\label{section:case}
Due to a large number of missed recalls found and limited human resources, we randomly sampled 8 representative missed recalls to discuss root causes with corresponding engineers. They accept all 8 missed recalls and summarize two major reasons causing missed recalls\footnote{ Due to privacy concerns, some characters in the shop name are replaced with *s.}. After discussion with corresponding engineers, \leak will be launched to scan for missed recalls for Meituan on a regular basis.
\subsection{Segmenting Related Missed Recalls}
General users construct natural language queries to express their search intentions. The search component first segments the query and decides which part conveys the main search intention. For some user queries like \textit{Fresh**de SPA}, where the two parts both carry important information about the user's intention, choosing a \textit{main part} might be tricky. For example, if the search component takes \textit{Fresh**de} as the part mainly carrying user search intentions, shops whose title contains the characters \textit{Fresh**de} could be recalled. In this case, as shops like \textit{Fresh**de Laundry} and \textit{Fresh**de Massage} exist nearby, the shop \textit{Fresh**de Healthy SPA} may be neglected. Hence, a missed recall occurs.
\subsection{Landmark Related Missed Recalls}
Some shop names contain locations. When a user wants to search for that shop on an e-commerce app, often a phrase referring to a location will be contained in the user query. When the search component gets this query, it must decide whether this location phrase specifies the location of the target shop or only states the name of the target shop. This judgment may lead to missed recalls. For example, there's a shop named \textit{F**'s Seafood Barbecue (Fangbang)}. When searching with \textit{Barbecue Fangbang}, the search component may loosely translate this query as 
\textit{find barbecue shops located at Fangbang}. As so many barbecue shops exist at Fangbang and only several of them can be presented to users, it is possible, although rare, that the shop \textit{F**'s Seafood Barbecue (Fangbang)} didn't get recalled. Hence, a missed recall occurs.
\section{Threats to Validity}\label{section:Thrests}

As \leak relies on GPT-3.5 turbo to generate test cases, stability of the service and hallucination issues may hinder the effectiveness of \leak. Firstly, incidents like network thrashing, API errors, and time-out errors can cause API calls to fail. To cope with this, we leverage the retry schema. If no content is returned after 30 seconds, we terminate the thread, wait a random period, and retry three times. 
Secondly, LLMs can not guarantee the reliability of the generated content due to hallucination. Thus, unreal test cases can be generated by \leak and introduce false positives. To cope with this, we customize a CoT prompt that mimics how general users construct user queries, leverage a \textit{rethink} step, and set the temperature to zero in implementation. 

Outside of the approach itself, the quality of input data affects the performance of \leak. Outdated input data as well as incorrectness in input data incur false positives. For example, if shops that no longer exist are taken as target shops, only false positives of missed recalls can be found.

\section{Related Work}\label{section:Related Work}
\textbf{Testing of Search Components.}
For user experience, search components are expected only to recall all relevant search results. Hence, precision and recall ~\cite{hawking2001measuring} have always been valued. However, relevance is considered a rather subjective concept~\cite{saracevic1975relevance}. As a result, search components were usually tested with human judgments ~\cite{gordon1999finding, su2003comprehensive} in the 2000s. For example, a good ranking strategy is expected to generate rankings similar to human-generated ones ~\cite{vaughan2004new}.
With the development of deep learning techniques, search components also leverage deep neuron networks to enhance their performances ~\cite{ganguly2015word, huang2020embedding, nigam2019semantic, zhang2020towards, li2021embedding}. The explainability issues~\cite{arrieta2020explainable} of deep neuron networks contribute to the oracle problem and motivate the introduction of metamorphic testing. ~\cite{zhou2015metamorphic} summarizes five groups of metamorphic relations and applies them to general-purpose search engines. Metamorphic testing can also be used in domain-specific search components like academic search facilities ~\cite{de2019applying}, e-commerce search functionalities~\cite{nagai2018applying}, and code search engines ~\cite{ding2020metamorphic}. To conduct metamorphic testing, the key step is to construct a reliable metamorphic relation, which serves as the oracle during testing. ~\cite{segura2022automated} provides an automatic method for generating metamorphic relations, which helps to alleviate the labor of constructing metamorphic relations. 

\textbf{LLM and Testing.}
Generally, LLMs are pre-trained language models following the Transformer architecture ~\cite{vaswani2017attention}. They usually feature large-scale training corpus and massive parameters. To leverage those pre-trained LLMs effectively, a group of researchers fine-tuned the models with downstream datasets ~\cite{devlin2018bert} before using them. LLMs have far more parameters than ordinary pre-trained language models, drastically increasing fine-tuning costs. To cope with this challenge, a series of cost-efficient strategies for fine-tuning are proposed, such as Lora~\cite{hu2021lora} and prefix tuning ~\cite{li2021prefix}. However, fine-tuning LLMs needs a large quantity of downstream data, which may not be accessible. Hence, in-context learning is an alternative ~\cite{wei2022chain}. Empirical studies are also conducted towards constructing more efficient prompts during in-context learning, such as ~\cite{gao2023constructing}. Due to their superior performances, LLMs are applied to multiple fields of research~\cite{biswas2023potential, jiao2023chatgpt, biswas2023role,pearce2023examining, fan2023automated, ahmed2023recommending, xia2023automated}, including tasks of software engineering~\cite{glass2002research}. 

In testing, LLMs can be leveraged in test case generation. TitanFuzz~\cite{deng2023large} uses a generative LLM to produce initial seed programs from target APIs and conduct mutation with them by LLMs to generate test cases for deep learning libraries. As TitanFuzz tends to generate ordinary programs, but edge cases generally induce more bugs, FuzzGPT~\cite{deng2023large} is proposed to learn from historical bug-triggering programs and generate similar test cases by LLMs. Bug reports are also effective resources for LLMs to generate bug-triggering programs~\cite{kang2023large}. Apart from generating test cases directly, LLMs can also guide the generating process to escape the coverage plateaus~\cite{lemieux2023codamosa}.

\textbf{Metamorphic Testing.}
Metamorphic testing~\cite{chen2003fault, liu2013effectively} is typically applied to untestable situations where a direct testing oracle is challenging to acquire. The core idea is to detect violations of metamorphic relations among multiple outputs~\cite{chen2018metamorphic, segura2016survey}. Many practices of metamorphic testing can be witnessed in the domain of software engineering~\cite{chan2005towards, lidbury2015many, mansur2021metamorphic}. Viewed as black boxes, AI-based systems also promote the application of metamorphic testing. It can be used to both classify algorithms~\cite{murphy2008properties, xie2011testing}, and software built upon them~\cite{zhang2018deeproad, tian2018deeptest, yu2022automated, wang2023mttm}. For demonstration, we take machine translation software as an example. Machine translation software typically leverages deep neuron networks to transfer one natural language to another. As the subjectivity of human expression, multiple sentences expressing the same semantics exist. Hence, there is no standard answer for a translation that can be used as the testing oracle. However, it can be observed that normally, sentences conveying different meanings should not have the exact translation, and translation of similar-structured sentences should typically exhibit the same sentence structure~\cite{gupta2020machine, he2020structure}. Hence, metamorphic relations could be established.

\section{Conclusion}\label{section:conclusion}
In this paper, we present \leak, the first testing approach targeting missed recalls of e-commerce search components.
\leak leverages an LLM, specifically GPT-3.5 turbo and a customized CoT prompt inspired by the historical missed recalls, to mimic how general users construct queries during online shopping and it relies on a metamorphic relation to find missed recalls.
Experiments with open data demonstrate its performance advantages to baselines. Experiments with private industrial data show that \leak discovers 101 confirmed missed recalls and only reports 17 false positives in the real industrial setting. Over half of the missed recalls which \leak discovered can be reproduced.

\section*{acknowledgments}
This work is supported by Meituan and the Natural Science Foundation of Shanghai (Project No. 22ZR1407900). 
We extend our heartfelt thanks to our colleagues in Meituan, specifically, You, Jingjian, Pingping, Xiaolan, Ying, and Qinling, for their kind help and support in
this work.
Y. Zhou is the corresponding author.
\balance
\bibliographystyle{ACM-Reference-Format}
\bibliography{lDetector/reference_new}

\end{document}